\documentclass[a4paper,12pt,oneside]{article}
\usepackage{setspace}
\usepackage{lineno}

\usepackage{fancyhdr}
\usepackage{textcomp}
\usepackage{color,soul}
\usepackage{hvfloat}
\newsavebox{\myimage}
\usepackage[utf8]{inputenc}
\usepackage[english]{babel}
 \usepackage[L7x]{fontenc}
\usepackage[authoryear]{natbib}
 \usepackage{mathtools}
 \usepackage[nottoc]{tocbibind}
 \usepackage{gensymb}
  \usepackage{url}
 \usepackage{afterpage}
 \usepackage[unicode=true, bookmarks = true]{hyperref}
 \usepackage[all]{hypcap}
 \usepackage{tocloft}
 \usepackage{anysize}
 \usepackage{fixltx2e}
 \usepackage{amssymb}
 \usepackage{caption}
 \usepackage{subcaption}
 \usepackage{setspace}
 \usepackage{amsmath}
\usepackage{array}
 \marginsize{3cm}{1cm}{2cm}{2cm}
 \hypersetup{
     colorlinks,
     citecolor=black,  
     filecolor=black,  
     linkcolor=red,   
     urlcolor=black    
 }


\begin{document}

 {\hypersetup{linkcolor=black}

  }

\noindent
\textbf{Low crater frequencies and low model ages in lunar maria: Recent endogenic activity or degradation effects?}

\noindent
A. Valantinas$^{1*}$, K. M. Kinch$^{1}$, A. Bridžius$^{2,3}$ 

\noindent
$^{1}$Niels Bohr Institute, University of Copenhagen, Øster Voldgade 5-7, 1350 Copenhagen, Denmark.
\noindent
$^{2}$Vilnius University Observatory, Čiurlionio 29, Vilnius LT-03100, Lithuania. 
\noindent
$^{3}$Center for Physical Sciences and Technology, Saulėtekio av. 3, Vilnius LT-10257, Lithuania

\noindent
*Corresponding author's e-mail: adomas.valantinas@gmail.com

  \newpage
  
  \section*{Abstract}
  
    Recently a number of studies have identified small lunar geologic structures to be <100 Ma in age using standard remote sensing techniques. Here we present new crater size frequency distributions (CSFD) and model ages using craters D> 10 m for 5 small target units: 1 Irregular Mare Patch (IMP) in Mare Nubium and 4 regions located on lunar wrinkle ridges in Mare Humorum. For comparison we also date another IMP found in a recent study in Mare Tranquillitatis \citep{braden2014}. Absolute model age derivation corresponds to 46$\pm$5 Ma and 22$\pm$1 Ma for Nubium and Sosigenes IMP. We show that for IMPs and in nearby control mare regions similar production-like cumulative log-log SFD slopes of -3 are observed. In contrast control mare regions in Mare Humorum exhibit shallower equilibrium slopes from -1.83 to -2. 3 out of 4 wrinkle ridges appear to be in equilibrium but with crater life times lower than on the corresponding maria. Low crater frequencies on one wrinkle ridge result in an age of 8.6$\pm$1 Ma. This study region contains 80\% fresh craters which suggests that the crater population is still in production indicative of a recent resurfacing event.

  \newpage
\section{Introduction}

  Crater size frequency distribution (CSFD) measurements provide a remote sensing tool for analyzing ages of planetary surfaces. This method was first devised and developed over several decades when radiometrically dated samples from the Apollo and Luna missions were correlated with observed lunar crater frequencies \citep{hartmann1970,neukum1983,neukum1994,neukum2001a}. It allowed researchers to estimate that the bulk of lunar maria basaltic plains formed from 3.8 Ga to 3.1 Ga ago \citep{basalts} while some examples exist of surfaces formed as recently as 1 Ga ago \citep{spudis1983,hiesinger2011}. By general knowledge active volcanism on the Moon ended at least one billion years ago. Recently many irregular structures of sizes ranging from a few kilometers to hundreds of meters have been found in meter-scale resolution images and their CSFDs measured, suggesting that their age of formation is within the last 100 Ma \citep{braden2014}.
  

  In this work we present new evidence for low crater frequencies and low model ages using small lunar crater (SLC) populations on the lunar surface. We applied CSFD measurements to 2 Irregular Mare Patches (IMPs) first identified by \citet{braden2014} in Mare Nubium (Nubium IMP) and in Mare Tranquilitatis (Sosigenes IMP). The latter was also dated in the work of \citet{braden2014}. We also analyzed 4 sections of 3 separate lunar wrinkle ridges in Mare Humorum. To our knowledge CSFD measurements have not been used to analyze lunar wrinkle ridges before. For comparison with the observed crater densities and derived model ages of these target regions we measured CSFDs on adjacent and similarly-sized control areas on the plains of Mare Tranquillitatis, Mare Nubium and Mare Humorum. Our analysis poses questions about the underlying surface processes which affect small lunar crater populations. Explanations for observed low counts of small craters in our work might point to global small scale resurfacing/degradation \citep{fasset2014,degradation} effects which also modify crater saturation equilibrium levels on these terrains \citep{werner2015}.
   
 Since our work relies on using small crater populations (D>10 m) to determine the extent of processes affecting small surfaces it is important to keep in mind some underlying assumptions of the crater count dating method, namely: 1) areas that were totally resurfaced and are accumulating craters exhibit a specific production SFD slope, 2) crater ages are less than their survival lifetime $T_{life}\sim2.5D(m)$ according to classic estimates of \cite{basilevsky1976}, and 3) the same scaling law is applicable to different terrains, i.e. the same impact makes the same size of crater on all surfaces.

\section{Background}

Currently, there is an ongoing discussion about when lunar endogenic activity stopped. Traces of radiogenic gases have been detected during past lunar missions \citep{benna} and the formation of the enigmatic structure named Ina \citep{schultz2006,ina} has been speculated to be linked to episodic gas release from deep within the Moon. More small irregular structures similar to Ina (IMPs) have been identified and cataloged by \citet{braden2014}. They are thought to have formed less than 100 Ma ago due to endogenic volcanic processes and hence, much later than all of the major lunar maria formation. Recent (< 50 Ma) tectonic activity has been proposed by \citet{watters2012} from observations of lunar graben on the farside highlands and mare basalts. Identifying geologic timescales for such structures is an important step in understanding lunar thermal evolution. Hence, CSFD measurements are often applied on analyzed structures to evaluate their model age.  

The method to derive model ages was originally devised for craters D> 1 km. However, today many CSFD age determinations have been performed on features of only a few square kilometers in extent such as impact melt pools, lobate scarps, ejecta blankets and IMPs using craters from a few hundred meters down to just a few meters in diameter (e.g. \citealt{zanetti,clark2015,hiesinger2012,braden2014}). 

Great concern was raised over secondary craters because if they influence local parent populations it would make surfaces exhibit artificially old model ages therefore rendering age dating problematic (e.g. \citealt{mcewen}). On the other hand \citet{hartmann2007} argued that contamination due to secondaries is minor and production functions can be safely used. \citet{ivanov2006} showed that craters D< 100 m found on young surfaces (age< 100 Ma) are primaries. Recent light has been shed about secondary craters by \citet{splotches} who compared 'after' and 'before' LRO images to conclude a substantial amount of small secondaries is evident from their parent primaries. The validity of using small impact craters for age measurements was also discussed by \citet{xiao&strom}. Counting on small \textit{surface areas} imposes another issue - validity of the statistics. However, recent studies have claimed that areas as small as 1 km$^2$ can be used for age dating \citep{bogert}. Thus, making the case to use this technique on small scale structures. 

In this work we also apply this method to target areas of lunar wrinkle ridges. Lunar wrinkle ridges are linear or sinuous asymmetric topographic highs that appear concentric or radial in respect to maria centers \citep{golombek1986} and can be up to 20 km in width, 300 km in length and 0.5 km in relief \citep{sharpton1988}. There is still an ongoing debate about details of their origin and formation but it is accepted to be an aftermath of tectonic processes. A recent global map of lunar wrinkle ridges and parametrization of their properties confirmed that they are closely linked with lunar maria and revealed that a global stress field was involved in their formation \citep{yue2015}. CSFD measurements have never been done before on lunar wrinkle ridges but could help answer questions about their origin and ongoing modification processes.

 Recently there has been a few studies about crater degradation \citep{fasset2014,degradation} which might be attributed to 'sandblasting' from small meteorites over billions of years, geometric overlap so called 'cookie-cutting', burial by ejecta, impact-induced or tectonic seismic events,  thermal creep, etc. Degradation affects craters of all sizes but smaller craters will disappear faster than larger ones. \citet{fasset2014} showed that degradation rates for large craters of 800 m - 5 km in diameter are up to a factor of 10 slower in maria than in the highlands. On the other hand, a \textit{faster} degradation rate at maria plains than in the highlands is observed for small lunar craters (SLCs) (35 m - 250 m) at the Apollo 17 landing site in the Taurus Littrow Valley by \citet{degradation}, who hypothesized that degradation rates might be dependent on local conditions such as terrain strength and cohesiveness. It has been shown that target surface properties can directly influence CSFD and therefore derived model ages \citep{oberbeck,properties,properties2}.   
 
Ultimately crater degradation at small sizes relates to saturation/equilibrium effects. A crater population reaches equilibrium at a particular size when craters are being produced at the same rate at which they are destroyed \citep{trask1966,gault1970,hartmann1984}. Crater saturation equilibrium is still not fully understood but recent studies by \citep{fasset2016,werner2015} suggests that terrain properties can affect the shape of the equilibrium CSFD through their effect on removal rates at different cater diameters. It is important to consider equilibrium effects because if a certain population reaches equilibrium that surface can no longer accumulate craters and the observed crater frequency does not express the original formation age of the surface structure in question. This is especially important when analyzing small craters as these will generally reach equilibrium sooner.

       \section{Methods and Data}       
       
    Age measurements on all of our selected areas were done using the JMARS \citep{jmars} software package and the data analysis software CraterStats \citep{michaelcrater}. Within JMARS craters and their enclosing area shapes were manually marked and then extracted into CraterStats2 (version last updated on 2017-03-19). To represent CSFD plots we use the differential method \citep{michael} because it reveals information about individual data bins while in the cumulative method a specific bin represents a certain crater number at a given range \textit{D} and all larger ones. This might conceal possible resurfacing effects among a crater population or any other deviances in the crater production. Our crater counts were binned into 10 bins per decade. Each bin boundary is 10$^{0.1}$ times the previous one. For example: 10, 12.6, 15.8, 20.0, 25.1, 31.6, 39.8, 50.1, 63.1 and 79.4 m. In some cases where the distribution fits an isochron over all or a sizable part of the size range, we extracted model ages by fitting the production function to data points in the relevant range. Age measurements were based on the \citet{neukum2001a} lunar chronology and production functions. We display all CSFDs on in R plot representation \citep{arvidson}. In those cases where we extracted model ages we also show differential plots. We compare crater distributions to the \citet{hartmann1984}  saturation equilibrium (HSE) which has a slope of -1.83 on a cumulative log-log plot. Also we use \citet{trask1966} saturation equilibrium (TSE) line which has a slope of -2. On a relative plot HSE has a slope of 0.17 and TSE a slope of 0. Historically R plots were devised to show the ratio of an observed crater population to the one with a cumulative slope of -2. Also, originally Hartmann's equilibrium curve was based on large craters and extrapolated to smaller diameters but we argue that some of our data are in good concordance with the cumulative slope of -1.83 in range D=10-100 m.             
          
    All crater identifications were done on Lunar Reconnaissance Orbiter (LRO) Narrow Angle Camera (NAC) images. A list of the NAC data frames used in this work is shown in Table~\ref{table:1}. Images of high solar incidence angle (66-72\degree) were used, due to distinct shadows which allow to distinguish craters from the background surface (e.g. \citealt{young1975}). Low resolution images were avoided where possible however due to incomplete and uneven high-res NAC coverage exceptions had to be made. Our image resolutions vary from 0.47-1.73 m/px. Identification of degraded craters around 10 m in diameter at resolution 1.7 m/px is challenging and might artificially lower counts at the very smallest diameters in some cases with coarser resolution.
   
\begin{figure}[h!]
               \centering
                \hspace*{-0.5cm}
               \includegraphics[width=15cm]{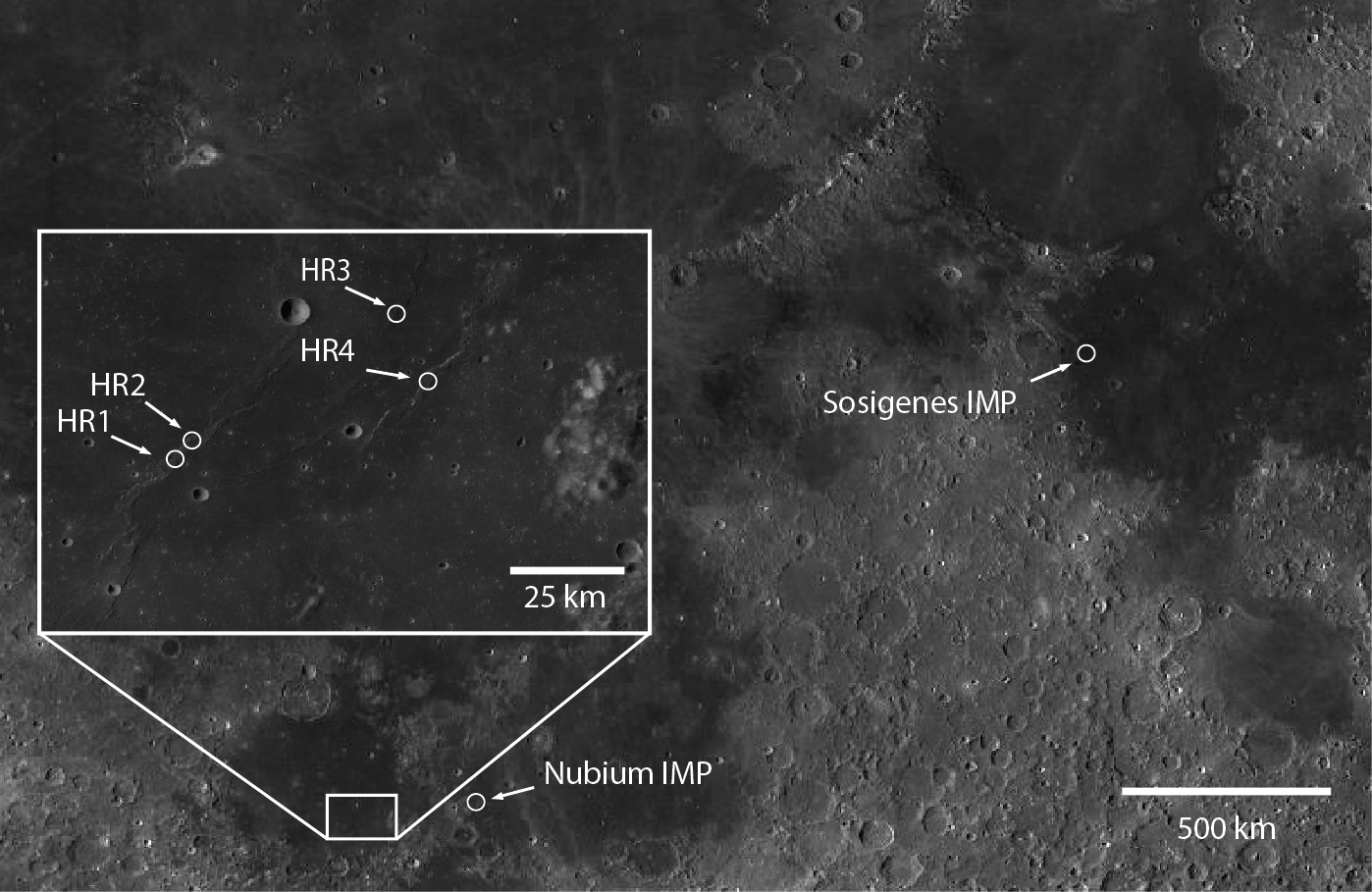}
                \caption{Global view of measured areas on the lunar nearside and a close up of analyzed region in Mare Humorum (Lunar Reconnaissance Orbiter Wide Angle Camera image). Sosigenes IMP found in Mare Tranquillitatis and Nubium IMP found in Mare Nubium shown with arrows. Zoomed in area includes areas HR1 to HR4 located on 3 wrinkle ridges on Mare Humorum.} 
                 \label{fig.global}
                 \end{figure}

\begin{figure}[h!]
                 \centering
                  \hspace*{-0.5cm}
                   \includegraphics[width=15cm]{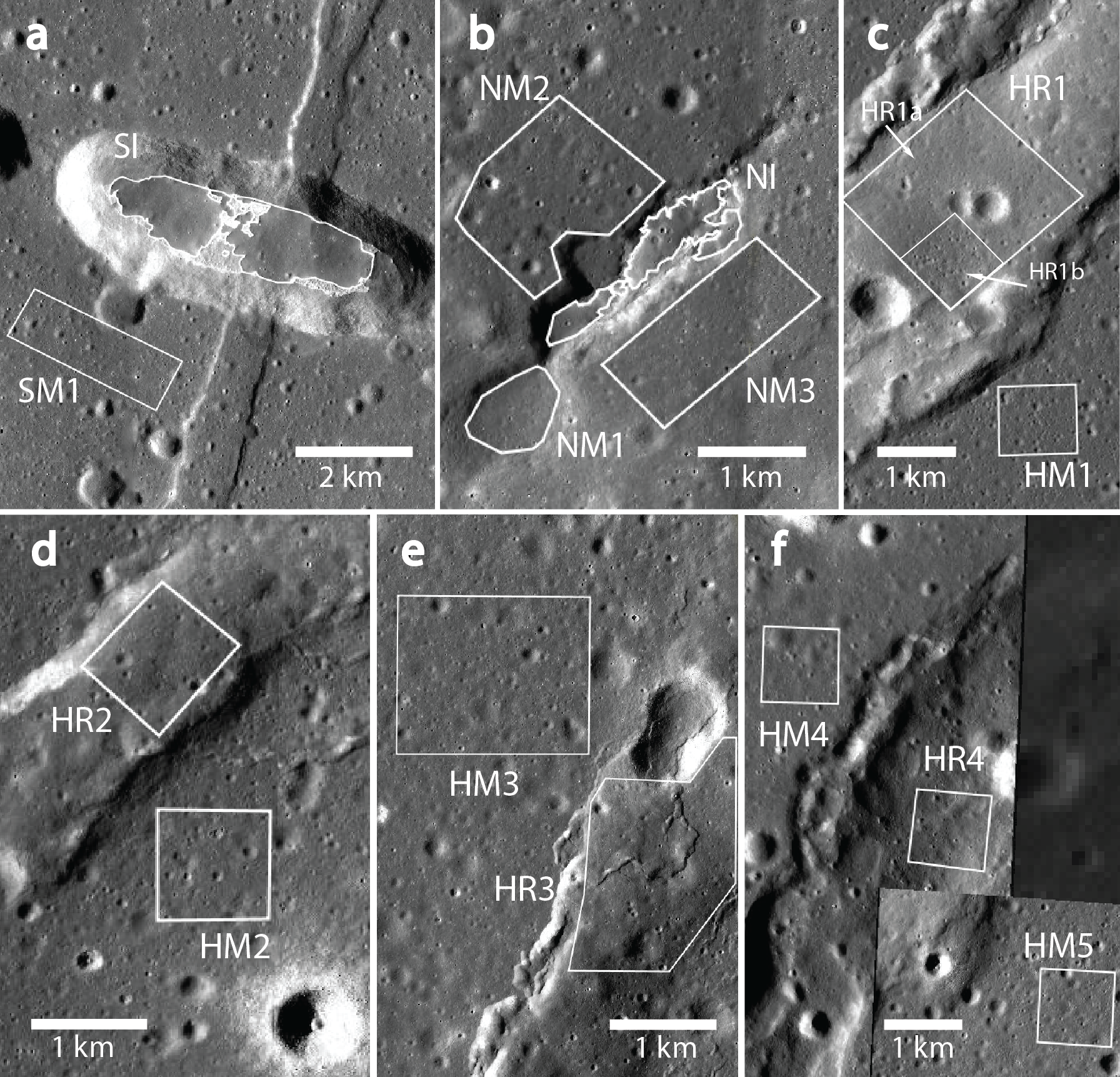}
                  \caption{Close up of areas used in CSFD measurements (Lunar Reconnaissance Orbiter Narrow Angle Camera images). Sosigenes IMP SI and control SM1 region (a). Nubium IMP NI and NM1-NM3 control count regions (b). Wrinkle ridges HR1-HR4 in Mare Humorum and their control areas HM1-HM5 (c-f). Image IDs can be seen in Table~\ref{table:1}. } 
                   \label{fig.countareas}
                  \end{figure}

    The location of all analyzed areas is shown in the overview map Fig. \ref{fig.global} and close-up images of each area are shown in Fig. \ref{fig.countareas}.   
                        
    We first analyzed 2 irregular mare patches (IMPs): Sosigenes IMP in Mare Tranquillitatis (Area designated \textbf{SI} located at: 8.35\degree N, 18.98\degree E, Fig~\ref{fig.countareas}a) and Nubium IMP in Mare Nubium (Area designated \textbf{NI} at -25.72\degree N, -27.67\degree E, Fig~\ref{fig.countareas}b). Sosigenes was already analyzed by \citep{braden2014} but we include it to compare to our results. Nubium was catalogued by \citet{braden2014} but its model age undetermined.                
                                        
    Next, we analyzed 4 areas on 3 separate wrinkle ridges in Mare Humorum: \textbf{HR1}(-27.17\degree N -37.39\degree E), \textbf{HR2} (-26.75\degree N, -36.89\degree E), \textbf{HR3} (-25.75\degree N, -35.30\degree E) and \textbf{HR4} (-26.35\degree N -35.02\degree E). HR1 was divided into 2 count areas: \textbf{HR1a} and \textbf{HR1b} to account for its local crater frequency differences. All wrinkle ridges showed smooth and homogeneous counting surfaces.The digital terrain model, GLD100 \citep{gld100} shows that they all exhibit low slopes (< 6\degree) and thus should not be affected significantly by mass wasting. 
                        
    For each of the IMPs and Wrinkle ridge areas we also analyzed nearby and similarly-sized control areas on the Mare plains (area \textbf{SM1} associated with Sosigenes IMP, areas \textbf{NM1-NM3} associated with Nubium IMP and areas \textbf{HM1-HM5} associated with the wrinkle ridges)
                                                             
    All regions were chosen to be at least 1 km$^2$ which was defined to be the smallest statistically robust area size for age dating using craters D>10 m \citep{bogert}. The exception is \textbf{NI}, which is only 0.71 km$^2$, limited by the small extent of the feature in question. Craters that had less than a half of their diameter in the counting area were excluded. All the areas we analyzed are listed in Table \ref{table:2} together with analysis results and auxiliary information.

   \section{Results}
      
 \subsection{Irregular mare patches}
                              
\begin{figure}[h!]
              \centering
              \hspace*{-0.5cm}
               \includegraphics[width=16cm]{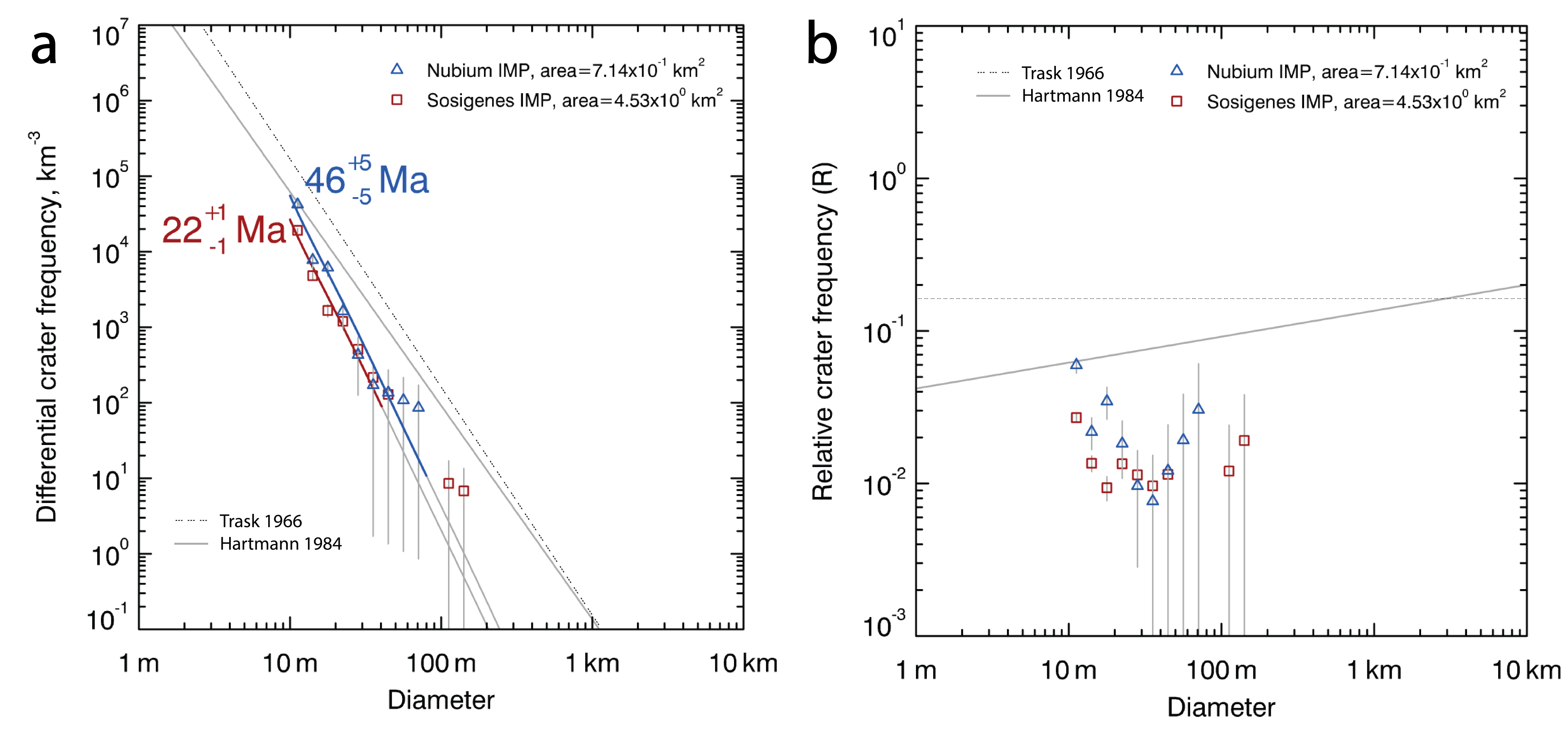}
              \caption{Differential Sosigenes and Nubium IMP CSFD plots (a). Absolute model age isochrons shown in red and blue. Relative crater frequency plot for the same study areas (b).} 
               \label{fig.NI_SI_rel}
               \end{figure}

	 The crater distribution for Sosigenes IMP (Fig. \ref{fig.NI_SI_rel}) exhibits a  slope significantly steeper than HSE or TSE and represents a model age of 22$\pm$1 Ma which is relatively close to the \cite{braden2014} result of 18$\pm$1 Ma. Only the 3 data bins at largest crater diameter show greater uncertainties and deviations from the production function due to low crater number per bin. The cumulative slope of the production function which is fit to the population of the Sosigenes IMP is close to -2.9. 
                                           
	Nubium IMP on the other hand, (Fig. \ref{fig.NI_SI_rel}) exhibits more scatter in the data points of the CSFD, which is likely related to the small areal extent of this feature (0.7 km$^2$). The fit to the production function gives a model age of 46$\pm$5 Ma. The crater diameter range 10-20 m is close to HSE, while larger crater diameter bins 20-40 m fall below HSE. 
                                                               
 \subsection{Wrinkle ridges}   
                
\begin{figure}[h!]
            \centering
            \hspace*{-0.5cm}
          \includegraphics[width=18cm]{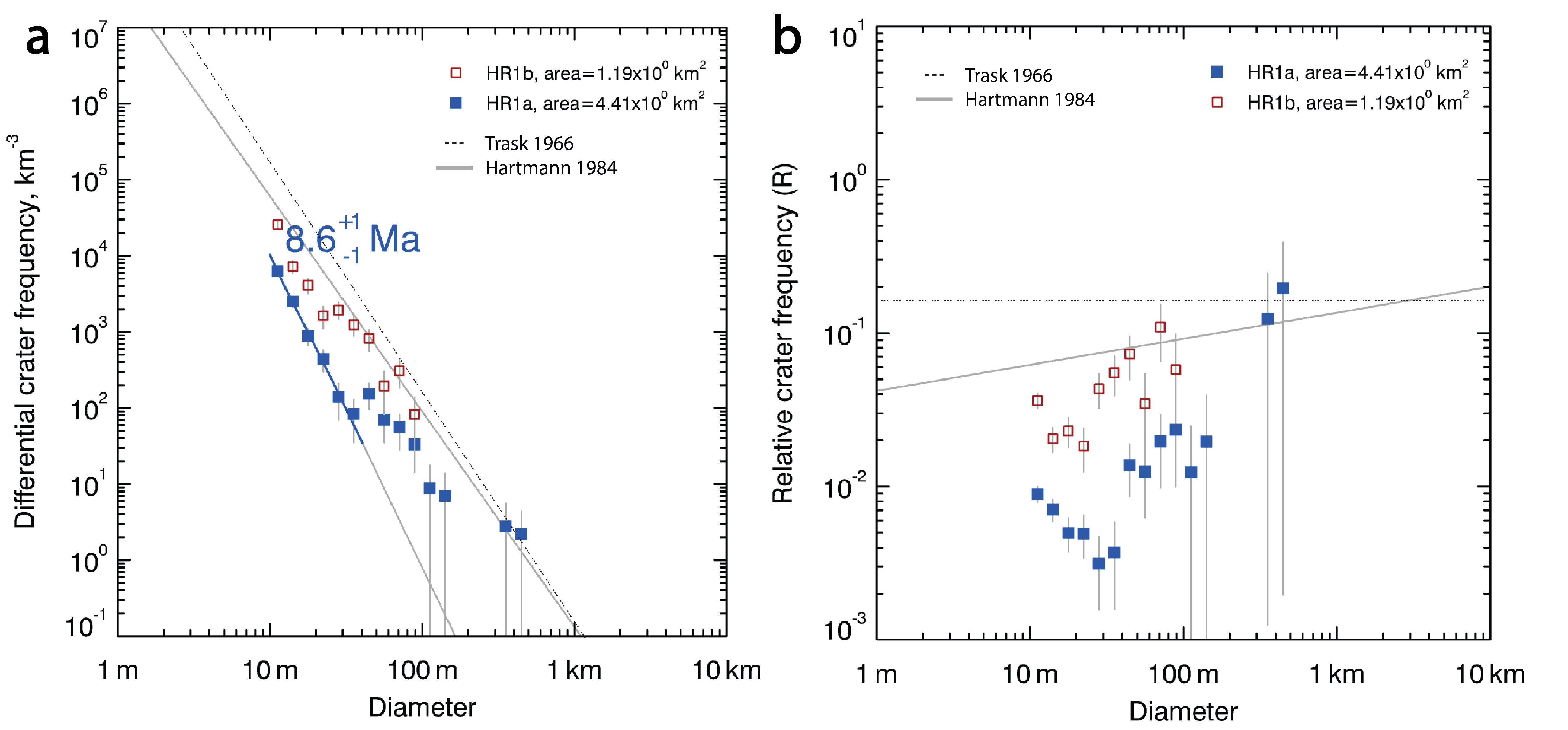}
        \caption{Differential and relative CSFD plots of areas HR1a (blue) and HR1b (red).} 
      \label{fig.hr1_both}
      \end{figure}

     HR1 exhibited the largest counting area among all analyzed regions (5.76 km$^2$). One can observe low crater frequencies even visually (Fig. \ref{fig.countareas}c). Most of its degraded craters are concentrated in the south western counting region and mostly small primarily fresh craters are present elsewhere. To account for this heterogeneity we split the counting area to HR1a and HR1b. From the CSFD plot (Fig. \ref{fig.hr1_both}b) we observe a difference in crater frequencies between HR1a and HR1b at D=10 of a factor of 4. Both HR1a and HR1b at D=~30 m exhibit a break off point between the smaller and larger craters which can be seen on a differential and relative plot (Fig. \ref{fig.hr1_both}a,b). The crater diameter range of area HR1a at 10-30 m resembles a slope of a population which is in production, with a model age of 8.6$\pm$1 Ma (Fig. \ref{fig.hr1_both}a). For area HR1b the range from 10-20 m resembles a production population and a fit in this range gives 32$\pm$4 Ma.
     
\begin{figure}[h!]
                       \centering
                  \hspace*{-0.5cm}
                \includegraphics[width=16cm]{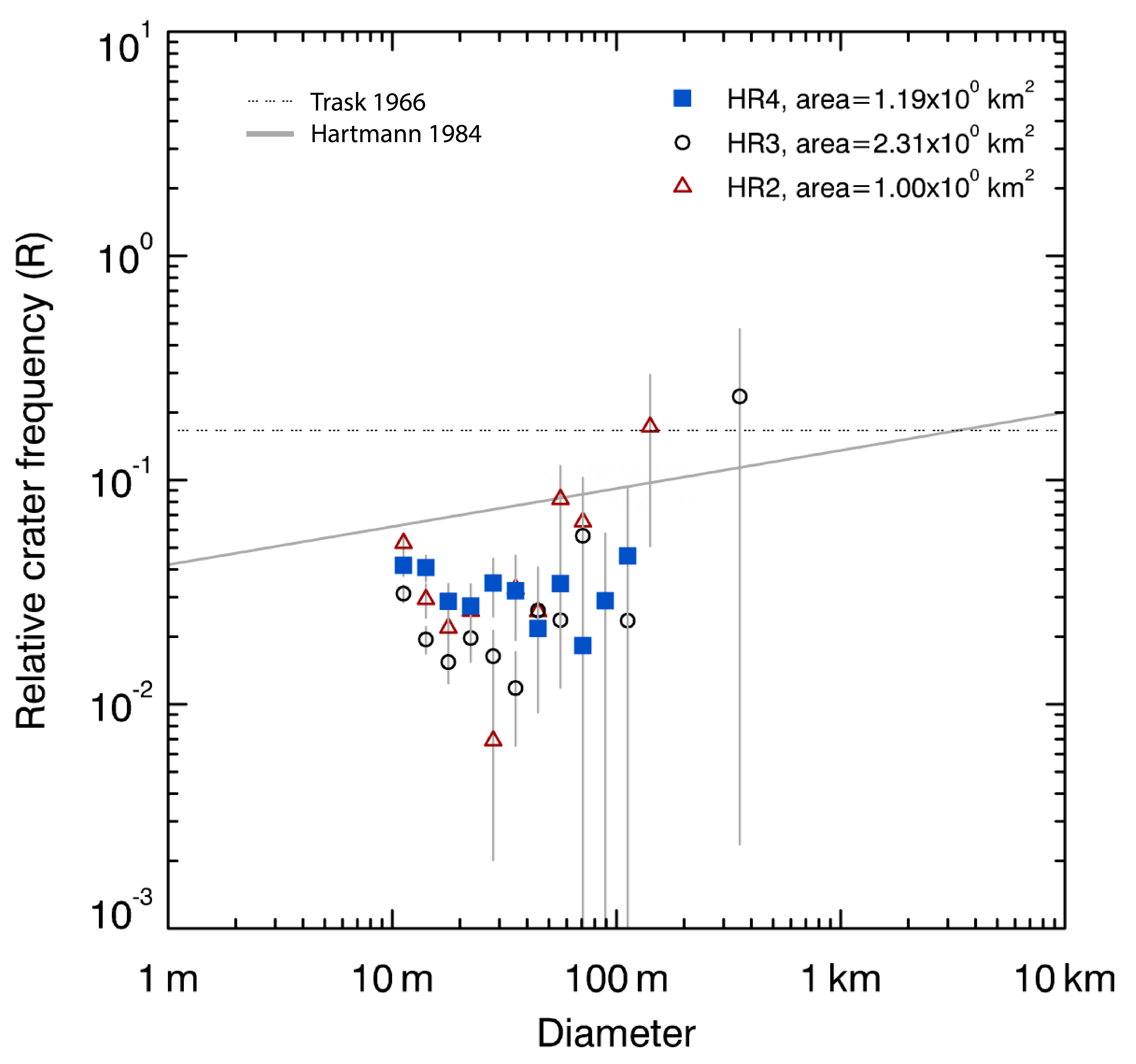}
                \caption{Relative CSFD plot for wrinkle ridge count areas HR2-HR4.} 
                 \label{fig.hr2-hr4_relative}
                 \end{figure}

     For areas HR2-HR4 we observe crater populations to be closer to HSE than HR1 (Fig. \ref{fig.hr2-hr4_relative}). The CSFDs also do not show signs of any particular slope. The break off point mentioned previously for HR1 is not evident for these surfaces between small and larger craters. They also fall far below the TSE.
     
    \subsection{Maria plains}   
  
\begin{figure}[h!]
           \centering
            \hspace*{-0.5cm}
           \includegraphics[width=16cm]{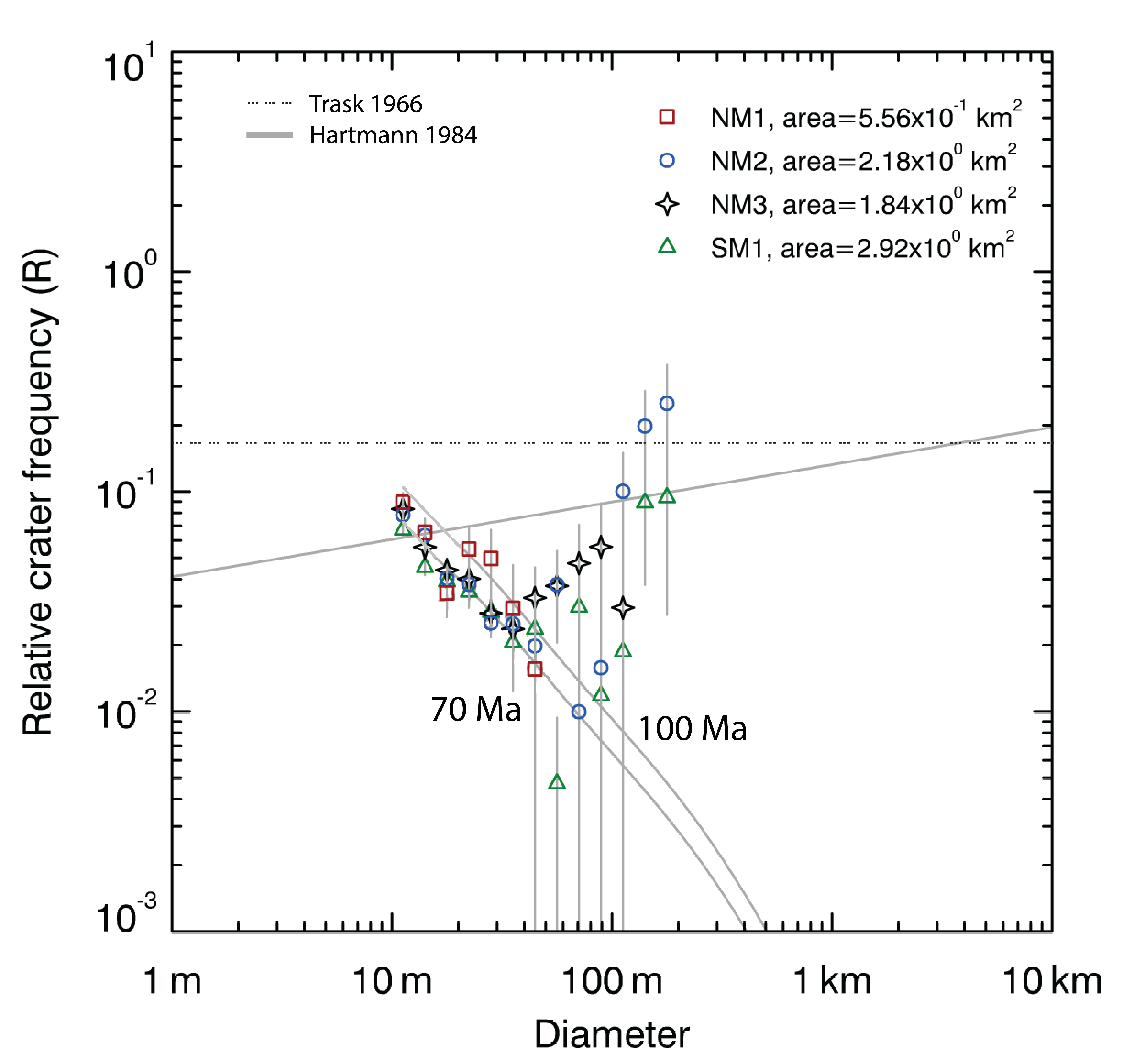}
           \caption{Relative CSFD plot of mare counts around IMPs in Mare Nubium (NM1-NM3) and Mare Serenitatis (SM1). A derived model age isochron for the observed crater population corresponding to 70 Ma and one representative of Tycho formation age -- 100 Ma.} 
            \label{fig.nm_sm_relative}
           \end{figure}

\begin{figure}[h!]
            \centering
            \hspace*{-0.5cm}
            \includegraphics[width=16cm]{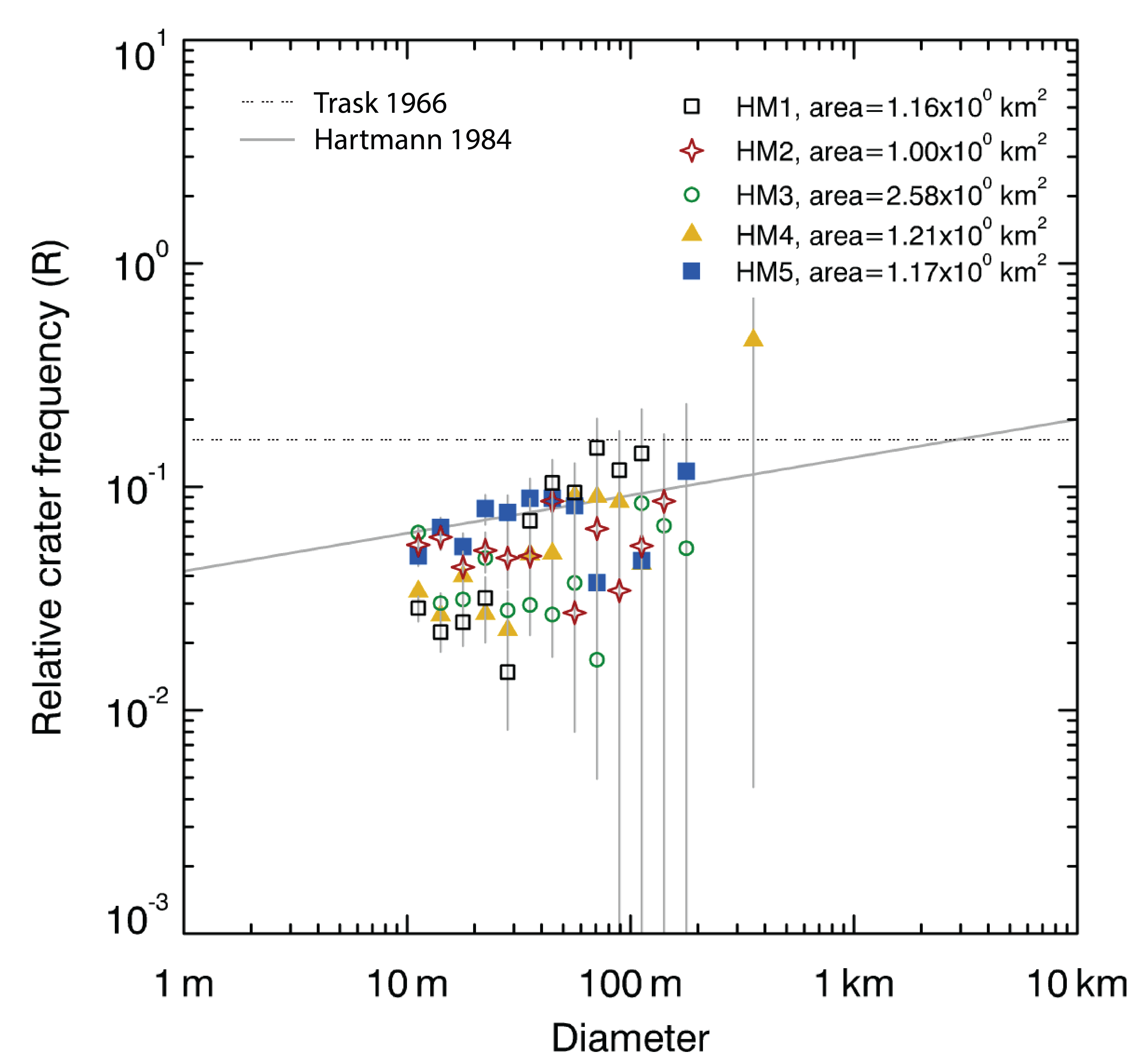}
          \caption{Relative CSFD plot of control mare regions in Mare Humorum (HM1-HM5).} 
          \label{fig.hm_relative}
            \end{figure}

    Analyzed crater count areas on relative frequency representation for maria plains in Mare Tranquillitatis, Mare Nubium can be seen in Fig. \ref{fig.nm_sm_relative}. Although areas SM1 and NM1-3 are located on different lunar maria we observe similar crater distributions. Steep distribution slopes for craters in range D=10-30 m are present for all count regions. This size range resembles a population in production. Around D>30 m larger craters do not follow the main distribution. This is somewhat similar to the break off point observed on HR1. These crater populations do not follow the HSE nor the -2 cumulative slope of TSE. In contrast, areas HM1-HM5 on Mare Humorum plains show (Fig. \ref{fig.hm_relative}) good agreement with HSE and TSE in terms of cumulative slopes closer to -1.83 or -2. Crater densities vary by a factor of 2 to 3 between different regions of Mare Humorum. We do not observe the 'V' shaped distribution as in Fig. \ref{fig.nm_sm_relative} nor the break off between larger and smaller craters. 
                
\subsection{Crater Morphology} 

\begin{figure}[h!]
                                        \centering
                                        \hspace*{-0.5cm}
                                        \includegraphics[width=15cm]{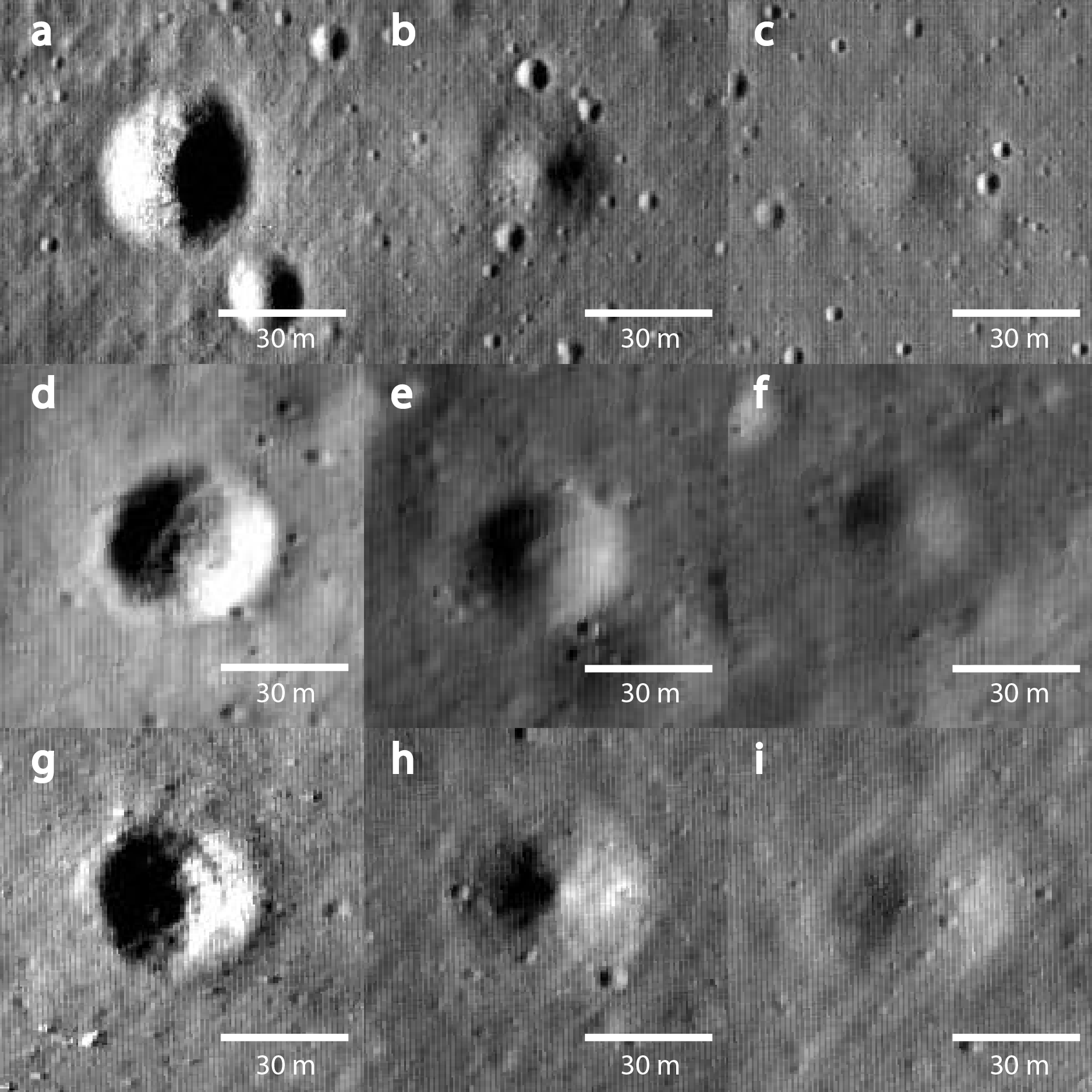}
                                        \caption{Crater morphology from 'fresh' to more degraded observed on Sosigenes IMP (a-c), wrinkle ridge HR2 (d-f) and Mare Humorum plains HM5 (g-i). Respectively: M177508146LE, M181408612LE, M10096730387RE.} 
                                        \label{fig.morphology}
                                        \end{figure}

   Within analyzed crater populations we observe a high variation in the morphology of craters. Even when images at high solar incidence  were used some craters were difficult to measure due to their vague physical features. It is important to mention that this was observed on IMPs, wrinkle ridges and maria plains universally. All analyzed areas exhibit craters in all stages of degradation. For example, in Fig. \ref{fig.morphology} three examples of craters from \textquoteleft fresh' to more degraded observed on Sosigenes IMP are shown, ones on one wrinkle ridge in Mare Humorum and craters observed in Mare Humorum plains.             
	
   We observed quite a variation in the percentages of different stages of degradation. For example on area HR1a in crater diameter range 10-30 m 80\% of craters appear to be fresh while in the D> 30 m 90\% of craters have degraded morphologies. This influences the steep slopes in range 10-30 m in Fig. \ref{fig.hr1_both}a which is a result primarily from fresh craters. For area HR1b the vast majority of craters in all diameter bins are degraded. The slope of HR1b (Fig. \ref{fig.hr1_both}a) at 10-20 m consists of ~20\% fresh craters. 

\section{Discussion}
\subsection{Maria and wrinkle ridges}

	Observed crater frequencies on mare areas near Sosigenes IMP, Nubium IMP and Mare Humorum exhibit discrepant results. In the IMP maria control regions (Fig. \ref{fig.nm_sm_relative}) crater distributions for NM1-SM1 have much steeper slopes than HSE or TSE and resemble populations which are still in production at D= 10-40 m. A break off point between small and large craters is evident at diameters ~40 m. However, control regions in Mare Humorum show populations (Fig. \ref{fig.hm_relative}) which are more in uniform along the crater diameter range. It is evident that even in the same maria densities can vary by a factor 2 to 3 on a relative plot. One can also observe that these populations exhibit a slope which is in better agreement with the HSE than the TSE. These maria counts are also in good concordance with Sinus Medii \citep{gault1970} and Hartmann's 'calibration' maria counts \citep{hartmann2005}. It may point to not a simple power law equilibrium. For lunar highlands different equilibrium slopes have been predicted in the past by numerical and analytical models \citep{chapman1986,richardson2009}. Such differences in equilibrium conditions due to local terrain properties or changes in crater accumulation has been also speculated by \citet{werner2015}. 
		
	CSFDs on wrinkle ridges in Mare Humorum (Fig. \ref{fig.hr2-hr4_relative}) appear to be similar to mare counts in shape. However, each corresponding maria area close to the ridge counts include generally higher densities, e.g. HR4 distribution is lower than HM4. This trend is evident for all count areas. Since the equilibrium level depends on the life time for a given crater diameter, one might conclude that the life time on wrinkle ridges is less than in the surrounding mare. Target cohesiveness differences as proposed by \citet{oberbeck} could have a direct influence on the rate of crater degradation on these wrinkle ridges. Material cohesiveness also has influence over crater sizes and more cohesive targets correspond to smaller crater diameters. This is important because modern scaling laws incorporate a \textquoteleft universal' target strength component when converting from projectiles to craters \citep[e.g.][]{ivanov2001}. In our work dates derived from standard CSFD analysis rely on the latter assumption.
	
\begin{figure}[h!]
                                           \centering
                                           \hspace*{-0.5cm}
                                           \includegraphics[width=16cm]{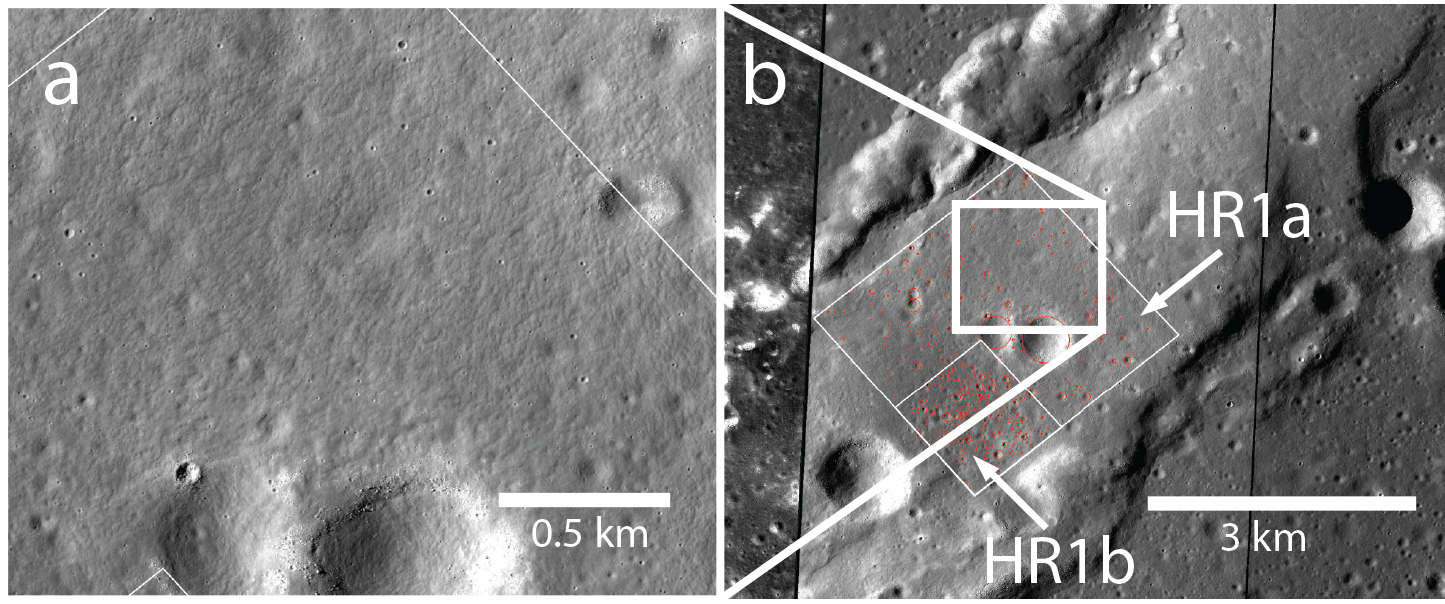}
                                           \caption{Wrinkle ridge count area HR1 (b) and a close up image (a). Lunar Reconnaissance Orbiter Narrow Angle Camera images.} 
                                           \label{fig.HR1}
                                           \end{figure}

	Count area HR1 is an anomaly if compared to other ridge counts. It is the largest area but has accumulated an unusually low number of craters (Fig. \ref{fig.HR1}). At the small end of the crater diameter range frequencies are lower than the ones observed in Tycho impact melt pools \citep{werner2015}. Craters from 10 m to 30 appear to be in production with a distinct slope for area HR1a (Fig. \ref{fig.hr1_both}). Due to the majority of craters found in HR1a being fresh, it is clearly not in an equilibrium state. HR1b on the other hand contains more degraded craters and is very close to equilibrium levels. The apparent split on CSFD plots from small (D=10-30 m) and larger (D>\textasciitilde40 m) craters of HR1a and HR1b areas are most pronounced among all studied wrinkle ridges. In the bins D= 10, 12.6, 15.8 20 and 25.1 m of count area HR1a there are 65, 28, 11, 5, 1 fresh craters vs 7, 8, 6, 5, 3 degraded ones. One can see that there are more fresh craters than degraded ones towards smaller sizes. Taking $T_{life}\sim2.5D(m)$ of \cite{basilevsky1976} one should expect a higher number of degraded craters in the smaller size bins because in the range of 10-20 m craters have survival times are 25-50 Ma. Since the resurfacing event 10 Ma ago a substantial amount of small craters should be degraded. This argues that either degradation is slower than given by Basilevsky on this surface or the majority of craters are even younger than 10 Ma. More degraded craters at large sizes seem to argue that some of them predated whatever event caused the removal of craters, but then the question is why we observe such a perfect production slope?
	
	 There are a few possible explanations for the resurfacing events responsible for low crater densities observed lunar wrinkle ridges. These tectonic structures which are in regions of contractional stresses could be affected by recent lunar moonquakes \citep{nakamura1980}. Other features known to have formed due to global contraction are lunar lobate scarps. They have also been identified to be young \citep[e.g.][]{binder1985,watters2010,clark2015}.

\subsection{IMPs}

	Our derived age of 22$\pm$1 Ma for Sosigenes confirms the young age derived by \citet{braden2014}. However, in this work we document 354 craters in D=10-25 m in comparison to Braden's 286. The difference in crater numbers might be associated with the various morphologies observed in these sites (Fig. \ref{fig.morphology}). A lot of craters on these surfaces were difficult to observe and measure due to their degraded characteristics which we believe is the major source of error. This could be improved by having better images in terms of resolution and incidence angles. Even though we observed \textasciitilde20\% more craters in our measurements (within expected uncertainty range as shown by \citealt{robbins2014a}), which corresponds to a slight vertically shifted model age isochron and therefore an older surface age, our derived age is still very young and thus consistent with the main result from Braden's work.
	
		Both Nubium and Sosigenes IMPs analyzed in this work show low crater densities with slopes much steeper than the equilibrium conditions found in Mare Humorum counts. However, in the small crater size range we observe similar production-like properties in the control IMP maria counts both in Mare Tranquillitatis and Mare Nubium (Fig. \ref{fig.nm_sm_relative}). For these areas we derived a model age of 70 Ma. According to \cite{basilevsky1976} the life time of craters in the range of 10-30 m is 25-75 Ma, so the production-like population of these craters should not exist or the life times of 10 m craters are 3 times larger than derived ones by Basilevsky. Due to crater degradation one would expect shallower slopes like in Mare Humorum control counts resulting in equilibrium (Fig. \ref{fig.hm_relative}). Possible explanations of these production-like slopes of mare counts around Sosigenes and Nubium IMP could be the following: 1) uneven and layered target material has nonuniform mechanical strength properties which requires a different scaling law, 2) contamination by secondaries in a narrow crater size range, and 3) complete resurfacing of these mare surfaces 70 Ma ago, implying that crater life time larger to a factor of 3 than proposed by \cite{basilevsky1976}. The first explanation seems unlikely that this surface was exactly \textquoteleft calibrated' to create a production-like slope from just gradual accumulation and degradation. On the other hand, the secondaries argument requires a removal of almost all craters from the original surface and the IMP to have significantly different target properties so the same impactor results in a smaller crater than on the mare. The third explanation is only possible if a massively longer crater lifetime than expected is observed and seems unlikely on a seemingly \textquoteleft normal' mare surface. Thus we argue that all of the three explanations are problematic. 				
	
	 Recently, the uniqueness of IMPs in comparison to lunar maria was investigated by \cite{elder2016} using the LRO Diviner instrument. The study found that the rock abundance on the analyzed IMPs is only slightly higher than the lunar average but much lower than the ejecta of ~100 Ma old craters that penetrate the regolith. It also reports no evidence for layering on analyzed surfaces and suggests the regolith to be at least 10 cm thick. Two hypothesized explanations are raised which implies that either the IMPs are older than observed crater distributions on IMPs or regolith is produced faster on lava flows than on blocky ejecta blankets. In this work, as mentioned previously, we observe very similar crater distribution slopes for Nubium and Sosigenes IMP as well as for neighboring mare control regions, namely SFD cumulative slopes close to the value -3 which is a characteristic of the craters being in production.

\subsection{Crater degradation}

	On all of our analyzed areas we see both fresh-looking craters and craters that appear to be in various stages of degradation (see Fig. \ref{fig.morphology}). Evolution of crater morphology over time which was investigated by several authors \citep[e.g.][]{basilevsky1976,fasset2014}. \citet{basilevsky2014} showed that in a typical mare equilibrium surface the percentage of craters with various depth to diameter ratios (the measurable degradation quantity) is the same for all diameter bins from 20 to ~200 m. In our work only the mare control areas in Mare Humorum support this observation. On the other hand, wrinkle ridge area HR1a (Fig. \ref{fig.HR1}b) contains in the diameter range 10-30 m 80\% fresh-looking craters versus HR1b with ~20\% fresh ones in the same size range. Larger diameters seem to be more dominated by degraded craters and for diameters above 60 m no craters appear entirely fresh on these surfaces. The high amount of fresh craters found at area HR1a mean that it hasn't yet reached the equilibrium condition \citep{chapman1970}. 
	
	According to recent study \citep{splotches} some of the morphologically degraded craters might be secondary impacts. This might make it difficult to distinguish between old degraded primary craters and more recent secondaries. However, in the CSFD plots secondaries form steep slopes and we do not observe this effect in Mare Humorum maria and wrinkle ridge counts. \citet{splotches} also showed that the top 20 cm of the lunar surface experiences full regolith gardening by small impacts in 10 Ma. According to our results at HR1 it seems that there are more dominant agents of crater degradation which are not gradual on timescales of million years but are rather sharp events able to garden up to a meter of regolith. The candidates for such degradation events could be nearby larger impacts redistributing regolith and/or seismic activity. However, we do not observe any ejecta rays or obvious sources of secondary cratering near studied regions. We also point out that the low slopes (\textasciitilde6\degree) on analyzed wrinkle ridges should have minimal effect on crater degradation. 
	
	 Crater degradation phenomena are closely tied with equilibrium processes. Within the crater studies community a few different equilibrium functions are in wide use without a unanimous agreement on a single model. The \citet{hartmann1984} (HSE) and \citet{trask1966} (TSE) functions have been used in this work. HSE was observed to be in better agreement with analyzed crater populations in Mare Humorum. In other literature \citep[e. g.][]{hiesinger2012} authors use the TSE function which has a cumulative slope of -2 and different coefficient factors than the HSE. TSE is also given along with HSE in CraterStats software \citep{michaelcrater}. If one were to compare measured CSFDs to each of these functions the observer might come to 2 different conclusions, e.g. \citet{hiesinger2012} only found 1 measured area in equilibrium according to the TSE function but the HSE implies that more surfaces would most likely have approached equilibrium. Recent work by \citet{fasset2016} and \citet{werner2015} sheds new light on equilibrium processes and indicates that no one universal function is applicable to all different lunar surfaces.

\section{Conclusions}

   In this work we present CSFD measurements based on LRO NAC images for 2 IMPs in Mare Nubium and Mare Tranquillitatis and 4 wrinkle ridges in Mare Humorum using small (D= 10-100 m) lunar crater populations on surfaces up to ~6 km$^2$ in spatial area. We also analyze 9 similar size nearby mare control regions to compare their crater distributions to the ones found on IMPs and wrinkle ridges. 
   
   Our results show that mare control regions around 2 well known IMPs in \citet{braden2014} study exhibit irregular CSFD cumulative log-log slopes of \textasciitilde-3 in contrast to general belief that equilibrated mare surface slopes are -2 \citep{gault1970}. Since these steep slopes are also observed on the IMPs it calls into question the idea of Nubium and Sosigenes IMP crater distributions being affected by a unique endogenic process. However, if absolute model age (AMA) derivation is applied on these IMPs we obtain 46$\pm$5 Ma for Nubium and 22$\pm$1 Ma for Sosigenes. The latter confirms the geologically young age and is in good agreement with 18$\pm$1 Ma found in \citet{braden2014} study.   
   
    Secondly, the identification of CSFDs on 4 wrinkle ridges in Mare Humorum reveal that crater life times are shorter than the nearby mare control regions. Crater densities appear to agree with Hartmann's maria calibration \citep{hartmann2005} and Sinus Medii counts \citep{gault1970}. However, it was observed that between the analyzed regions of Mare Humorum crater densities can vary by a factor of 2 to 3 and equilibrium cumulative slopes are in range from -1.83 \citep{hartmann1984} to -2 \citep{trask1966}. 
     
    One of the analyzed wrinkle ridges revealed very low crater densities (Fig. \ref{fig.HR1}) - lower than the ones found on IMPs by \citet{braden2014}. This region contains a majority of fresh craters which is an indication of a surface being in production, not equilibrium \citep{chapman1970}. The absolute model age derived for this unit corresponds to 8.6$\pm$1 Ma. Authors are not aware of a lower age ever recorded by previous workers for lunar surfaces. We suggest that this could be evidence for resurfacing by some discrete event which redistributes the top layer of regolith enough to erase all small craters entirely.
            
  \section{Acknowledgments}          
      Authors are thankful to reviewers B.A. Ivanov and C.R. Chapman for extensive and valuable comments which helped much to improve this work.  
          \newpage
 
      \section*{Tables \& Figures}
                   
                                    \begin{table}[!h]                               
                                  \centering               
                                  \caption{Details of images used for crater counts.}  
                                   \begin{tabular}{c c c c } 
                                 \hline 
                                  Image ID & Date & Resolution (m/px)& Incidence Angle (\degree)
                                  \\  
                                  \hline                  
                                    \hspace{3cm}     \textit{Sosigenes} 
                                    \\             
                                 M192824968RE       & 2012-05-28 & 1.182  & 70.86   \\                                            
                                 M192832116LE   & 2012-05-28 & 1.159  & 70.02    \\
                                       \hspace{3cm}    \textit{Nubium} \\
                                 M1142616950LE    & 2013-12-25 & 0.8   &  72  \\                            
                                 M1142616950RE  & 2013-12-25 & 0.8 & 72   \\ 
                                       \hspace{3cm}    \textit{HR1 \& HM1} \\
                                 M1173292388LE    &  2014-12-15 & 1.715   &  71  \\          
                                 \hspace{3cm}   \textit{HR2 \&HM2}    \\ 
                                 M181408612LE    & 2012-01-17 & 0.859   &  66.53  \\                                                                                                  
                                  \hspace{3cm} \textit{HR3 \& HM3} \\                 
                                 M1096730387RE    & 2012-07-12 & 0.65  &  69 \\
                                        \hspace{3cm}   \textit{HR4 \& HM4-5} \\
                                 M1154442219LE & 2014-05-11 & 1.829 &  72.83 \\
                            	 M1173278183RE & 2014-12-15 & 1.732 &  71.71 \\   
                            	                                                  
                                           \hline 
                                  \end{tabular}
                                  \label{table:1}
                                  \end{table}

                        \begin{table}[!h]
                        \small 
                        \caption{Unit name, size of counting areas, cumulative number of craters counted at different size range intervals, cumulative crater density at N(D=1 km) per square kilometer, absolute model ages (AMAs) and errors for each measured unit. No age dating derived for units close to equilibrium, shown in dashes.}                  
                         \begin{tabular}{c c >{\centering}m{2cm} >{\centering}m{2cm} >{\centering}m{2cm} >{\centering}m{2cm} c }
                        \hline 
                        Unit & Area (km$^2$)& \# of craters D$\geq$10 m    & \# of craters D$\geq$25 m & \# of craters D$\geq$50 m & N(1)($\times 10^{-5}$) & AMA (Ma) \\
                        \hline                              
                         SI  & 4.54 & 385  & 34  & 3 & 1.86 & 22$\pm$1  \\
                         
                         SM1  & 2.92 & 696  & 63  & 12 & \textminus & \textminus \\     
                         NI  & 0.71 & 126  &  6 & 2  & 3.87 & 46$\pm$5 \\   
                         NM1  & 0.56 & 173  &  14 & 0 & \textminus & \textminus \\ 
                         NM2  & 2.18 & 626  &  56 & 21 & \textminus & \textminus  \\
                         NM3  & 1.84 & 536  &  45 & 14 & \textminus & \textminus  \\  
                         HR1a  & 4.41 & 163  &  31 & 16 & 0.72 & 8.6$\pm$1 \\
                         HR1b  & 1.19 & 185  &  50 & 12 & 2.71 & 32$\pm$4\\                         
                         HM1  & 1.16 & 186  & 59 & 26 & \textminus & \textminus\\                
                         HR2  & 1.00 & 186  & 23  & 13 & \textminus & \textminus \\
                         HM2  & 1.00 & 267  & 43 & 9 & \textminus & \textminus \\  
                         HR3  & 2.31 & 266 & 39  & 15 & \textminus & \textminus\\
                         HM3  & 2.58 & 560   &  69 & 16 & \textminus & \textminus\\    
                         HR4  & 1.19  & 215  & 30  & 6 & \textminus & \textminus \\                                                
                         HM4  & 1.21 & 351  & 74 & 11 & \textminus & \textminus \\                                                                                                      
                         HM5  & 1.17  & 206  & 48  & 20 & \textminus & \textminus\\  
                              \hline          
                        \end{tabular}
                        \label{table:2}
                        \end{table} 
   
   \clearpage   
      \section{Supplement}                  
                                
\hvFloat[
 floatPos=!htb,
 capWidth=h,
 capPos=r,
 capAngle=90,
 objectAngle=90,
 capVPos=c,
 objectPos=c]{figure}{\includegraphics[width=25cm]{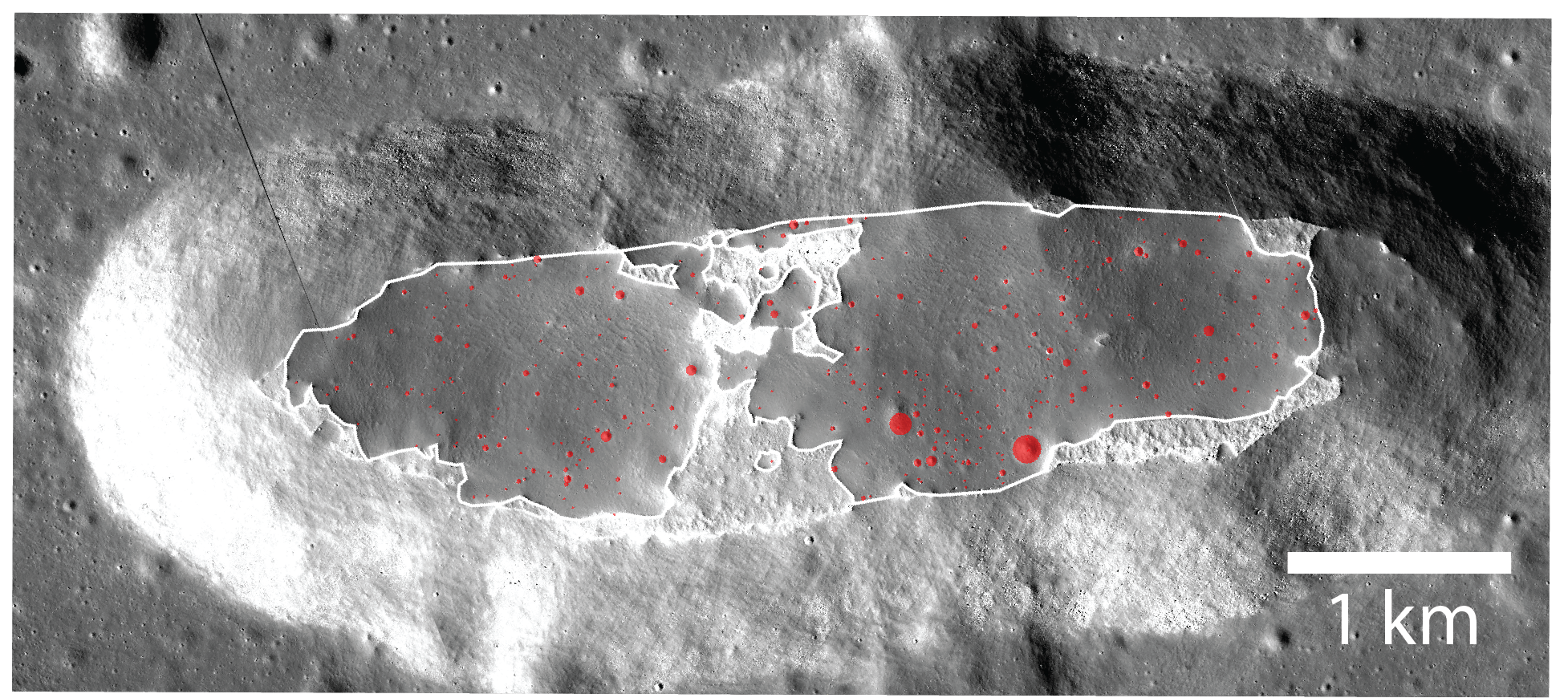}}%
{High resolution image of Sosigenes Irregular Mare patch (IMP) and its crater counts. Lunar Reconnaissance Orbiter Narrow Angle Camera image. Image IDs given in Table~\ref{table:1}.}{fig.sosigenes}

    \clearpage                    
\bibliographystyle{apalike}
 \renewcommand\bibname{References}
\bibliography{article_v3}

  \end{document}